\documentclass[runningheads]{llncs}

\usepackage{booktabs}
\usepackage{graphicx}
\usepackage{subcaption}
\usepackage{amssymb}      
\usepackage{pifont}       
\usepackage{url}

\begin{document}

\title{Before the Pull Request: Mining Multi-Agent Coordination}

\author{Dipankar Sarkar}
\institute{Arizona State University \\ \email{dsarkar3@asu.edu}}

\maketitle

\begin{abstract}
Autonomous coding agents now open millions of pull requests, yet large-scale studies find
their PRs are produced faster but accepted less often---a coordination and trust gap that
pull-request-level telemetry cannot explain. We argue the missing signal lives \emph{before} the
PR, in how concurrent agents claim, divide, and collide over shared work. We study this process
through \emph{grite}, our open-source coordination substrate that needs no central server and
stores its records inside git itself, so its append-only, signed event log captures the
coordination process directly. We show that (i)~this shared substrate reduces duplicate and
conflicting work at bounded overhead---the share of work that merely re-does a teammate's task
falls from $78\%$ to $0\%$ while useful throughput more than triples; (ii)~every agent's copy of
the log converges to the same state with no write silently dropped, where a file-based tracker
loses concurrent writes; and (iii)~the log is a mineable artefact from which concrete failure
modes---conflicting edits, lock starvation, redundant rediscovery, race-to-close---are
automatically recoverable with provenance, several invisible in pull-request history. We release
the dataset, harness, and mining toolkit.

\keywords{AI for software engineering \and multi-agent coordination \and
repository mining \and CRDTs \and human--AI collaboration \and coding agents}
\end{abstract}

\section{Introduction}
\label{sec:intro}

Autonomous coding agents have moved from autocomplete to teammates. Systems such as OpenAI
Codex, Devin, GitHub Copilot, Cursor, and Claude Code now open, review, and merge code at
scale: the AIDev dataset records over 456{,}000 pull requests authored by five such agents
across 61{,}000 repositories~\cite{li2025se3}. Yet the same large-scale analysis surfaces a
tension: agent pull requests are produced \emph{faster} than human ones but are accepted
\emph{less often}~\cite{li2025se3}.

Most explanations for this gap look \emph{inside} the pull request---code quality, test
coverage, reviewer load. We argue that an important part of the answer lies \emph{before} the
pull request, in a layer current datasets cannot see: the \emph{coordination process} by which
concurrent agents claim, divide, and collide over shared work. When several agents operate on one
codebase, two may pick the same task, edit the same issue, or duplicate a fix a teammate already
landed. None of this survives in commit or PR history---an abandoned duplicate never becomes a
PR; a task two agents raced to close leaves only the winner's trace---so the process that produces
redundant and conflicting work is invisible precisely where we most need to study it.

To study that process we built \emph{grite}, our own open-source coordination substrate for AI
agents (Section~\ref{sec:design}); this paper makes grite both the object and the instrument of
study. It is \emph{server-less}---there is no central coordination service; agents coordinate by
reading and writing shared task records---and \emph{git-native}---those records live inside git
refs, not a working-tree file or external database, so they travel with the code on ordinary
\texttt{git fetch} and \texttt{push}. Each coordination action is one entry in an append-only,
content-addressed, optionally signed event log, and per-agent copies are reconciled with
conflict-free replicated data type (CRDT) semantics~\cite{shapiro2011crdt} plus advisory leases
for mutual exclusion. This gives the paper two things at once: because coordination state is
shared and conflict-free we can \emph{measure} how the substrate changes outcomes, and because
every action is a typed, provenance-bearing event the log is itself a \emph{mineable
software-engineering artefact}---the pre-PR telemetry that PR-outcome datasets lack.

\paragraph{Contributions.} The paper contributes (1)~\emph{grite}, our open-source server-less
git-native coordination substrate for concurrent coding agents; (2)~a controlled, reproducible
measurement of how such a substrate changes coordination outcomes; and (3)~a mineable pre-PR
coordination dataset and mining toolkit that recover failure modes invisible to PR-outcome
datasets. We frame these as three claims, each backed by one experiment.
\textbf{C1 (coordination efficiency):} advisory leases plus shared task state reduce duplicate
and conflicting work at bounded overhead---the duplicate-work rate (the share of completed work
that merely re-does a teammate's task) falls from $0.78$ to $0.00$ while goodput (distinct tasks
per round) more than triples. \textbf{C2 (convergence without data loss):} replicas receiving the
same events in any order converge to byte-identical state, and concurrent writes are preserved
where a file-based tracker silently loses them. \textbf{C3 (a mineable process):} the log admits
automatic detection of concrete failure modes---conflicting edits, redundant rediscovery, lock
starvation, race-to-close---with provenance, several unrecoverable from PR history; mining also
shows advisory leases \emph{alone} do not prevent redundant rediscovery, whereas leases plus
shared state do. We release the dataset, harness, and toolkit, reproducible from a pinned commit
and fixed seeds.

\section{Background and Related Work}
\label{sec:background}

grite (Section~\ref{sec:design}) stores an agent issue tracker as an append-only event log in git
refs, rebuilds issue state by CRDT merge, and adds advisory leases for mutual exclusion. We
position it against four lines of work.

\paragraph{Mining what agents produce.}
Empirical SE increasingly mines the output of agents and bots. AIDev characterises hundreds of
thousands of agent pull requests and reports the speed--acceptance gap that motivates this
paper~\cite{li2025se3}. It builds on a longer line of mining-software-repositories work that
identifies and studies bots committing code~\cite{dey2020bots}. Both operate on
\emph{outcomes}---merged or rejected PRs, landed commits. Neither can observe the coordination
that happens \emph{before} a PR exists, which is the process we study.

\paragraph{Agent memory and multi-agent frameworks.}
Most agent ``memory'' today is retrieval. The Model Context Protocol exposes external stores to
an agent~\cite{mcp2024}, and retrieval-augmented generation conditions an agent on fetched
context~\cite{lewis2020rag}. These provide recall, not coordination: they offer no mutual
exclusion and no shared, provenance-bearing write history. Multi-agent frameworks such as
AutoGen orchestrate several agents within one process or session~\cite{wu2023autogen}, which
assumes a shared runtime rather than decentralised agents that synchronise through a repository.
Single-agent SE benchmarks and agents---SWE-bench~\cite{jimenez2024swebench} and
SWE-agent~\cite{yang2024sweagent}---evaluate task resolution by one agent, not coordination
between many.

\paragraph{Git-native and repository-embedded trackers.}
Embedding issues in the repository itself is established practice. Fossil keeps tickets in the
same versioned store as the code~\cite{fossil}; git-bug stores issues as native git
objects~\cite{gitbug}; and Beads is a recent git-backed dependency-graph tracker aimed at
agents~\cite{yegge2025beads}. These share grite's offline-first stance, and Beads is the closest
comparison. They differ in their reconciliation model: none combines a formal CRDT projection of
issue state, advisory leases for mutual exclusion, and a signed, content-addressed log in one
substrate.

\paragraph{Consistency and integrity foundations.}
grite composes well-understood building blocks. CRDTs give strong eventual consistency without a
coordinating server~\cite{shapiro2011crdt}, in contrast to operational transformation, which
historically relied on a central server to order edits~\cite{ellis1989concurrency}; local-first
principles motivate grite's offline-first design~\cite{kleppmann2019localfirst}. For mutual
exclusion, lease services such as Chubby require a consensus protocol~\cite{burrows2006chubby};
grite instead uses advisory leases over git refs and obtains convergence from CRDT merge, so it
needs no server and no agreement round.

\noindent Two gaps remain. First, no prior system unifies conflict-free concurrent agent edits,
advisory leases, and a signed, content-addressed, mineable history in a server-less git substrate:
server-based trackers (e.g.\ GitHub Issues) are neither offline nor conflict-free, file-based git
trackers lack a formal CRDT and leases, and retrieval memory offers neither mutual exclusion nor a
provenance-bearing log. Second, no prior dataset exposes the pre-PR agent coordination process for
empirical study.

\section{grite: A Git-Native Coordination Substrate}
\label{sec:design}

grite represents an issue tracker as an append-only event log living in git refs
(\texttt{refs/grite/wal}), with a materialised view (an embedded key--value store) rebuilt from
that log for fast queries. Nothing is written to the working tree, so coordination state travels
with the code through ordinary \texttt{git fetch} and \texttt{push}. We summarise the four
mechanisms below. Each populates fields of the exported coordination log---one row per event,
with fields \texttt{event\_id}, \texttt{actor\_id}, \texttt{ts\_ms}, \texttt{issue\_id},
\texttt{kind}, \texttt{conflict}, \texttt{duplicate}, and \texttt{lock\_outcome}. For example, a
\texttt{state\_changed} event by actor \texttt{0e} on issue \texttt{6} with \texttt{conflict=1}
(a cross-actor overwrite) and \texttt{lock\_outcome=denied}. The full schema is in
\texttt{data/SCHEMA.md}; our three claims each rest on a subset of these fields, noted with each
mechanism.

\paragraph{Typed, content-addressed, signed events.}
Every coordination action is an event with a kind (issue created/updated, comment, label
add/remove, state change, dependency add/remove, and others), an actor identifier, and a
millisecond timestamp. The event identifier is a BLAKE2b hash of its canonical encoding, so any
tampering invalidates the id, and events may additionally be signed (Ed25519). This is what makes
the log a mineable artefact with verifiable origin rather than a mutable database, and the basis
for the provenance the mining toolkit reports (C3).

\paragraph{CRDT projection.}
The materialised state of an issue is a projection over its events. Scalar fields (title, body,
state) use last-writer-wins keyed on the total order $(\textit{timestamp}, \textit{actor},
\textit{event\_id})$; sets (labels, assignees, dependencies) are commutative; comments and links
are append-only. Because the rebuild applies events in this canonical order, two replicas that
have seen the same events compute identical state regardless of delivery order. This is the
convergence property behind claim~C2. We instrument the projection to record, per applied event,
whether it resolved a \emph{cross-actor} conflict (a last-writer-wins overwrite of, or by, a
different actor's value); this records the \texttt{conflict} field and
is the conflicting-edit signal behind C1 and C3.

\paragraph{Advisory leases.}
Agents coordinate exclusive work through TTL-bounded leases stored under
\texttt{refs/grite/locks}. A lease is acquired before working a resource, renewed while work
continues, and released on completion; expiry bounds the damage of a crashed or stalled agent.
Each acquire, renew, release, expiry, or denial is recorded in the \texttt{lock\_outcome} field,
so the lease stream is itself mineable; denied acquisitions are the starvation/overhead signal
behind C1 and C3. Leases are \emph{advisory}---an agent may ignore one---which is itself a
measurable behaviour we return to when discussing partial compliance.

\paragraph{Dependency graph and sync.}
Issues carry typed edges (\textit{blocks}, \textit{depends\_on}, \textit{related\_to}) with cycle
detection, letting agents plan ordered work. Synchronisation is a plain fetch/push of the grite
refs followed by a CRDT merge; the design is offline-first, with no central server and no
consensus round (contrast Chubby~\cite{burrows2006chubby}). The git WAL is the source of truth,
the CRDT projection a materialised view over it.

\section{Experimental Methodology}
\label{sec:methodology}

\paragraph{Independent variable: the coordination arm.}
The one variable we manipulate is the \emph{coordination arm}---how much coordination machinery
the agents are given. We compare three. Under \emph{no-coord}, agents pick tasks freely; nothing
prevents two from working, and re-completing, the same task (the default for independent agents
today). Under \emph{locks-only}, an agent takes an exclusive advisory lease before working a task,
so no two agents work it at the same instant, but there is no shared record of what is already
done. Under \emph{locks+state}, agents additionally consult shared task state and skip a task a
teammate has already completed.

\paragraph{Agents and the task pool.}
For clean causal claims we use deterministic, seeded \emph{tier-T1} agents rather than LLMs. The
task pool is an abstract set of independent work units---not real source files---which lets us
control contention precisely and removes code-quality confounds; several tasks overlap, so more
than one agent may select the same one. Each agent repeats a loop: select a candidate task; in
the lease arms, try to acquire its lease and back off on denial; then ``work'' it by emitting
real grite events (\texttt{select}, \texttt{issue\_updated}, \texttt{state\_changed} to closed)
through the instrumented CRDT projection. Because these are genuine events on grite's data model,
the conflict and duplicate flags are computed by the substrate, not modelled. We sweep
$N \in \{2,4,8,16,32\}$ over seeds with pool size fixed, so contention rises with $N$.

\paragraph{Dependent variables.}
We report four metrics, each computed directly from the event log. \emph{Duplicate-work rate} is
completions of an already-completed task over total completions---the fraction of finishing work
that re-does a teammate's task. \emph{Conflicting edits} is the count of cross-actor
last-writer-wins overwrites (events whose \texttt{conflict} flag is set by \texttt{apply\_tracked}).
\emph{Goodput} is distinct tasks completed per round. \emph{Lock denials} is the number of denied
lease acquisitions, a proxy for coordination overhead and starvation.

\paragraph{Dataset and path to real agents.}
The analysis runs on the tidy, one-row-per-event coordination log
(\texttt{data/coordination-log.csv}; schema in \texttt{data/SCHEMA.md}), emitted directly by the
harness. All quantitative results here are tier-T1 (synthetic). The same detectors in
\texttt{mine/} are agnostic to the log's origin: \texttt{grite export --format coordination-log}
flattens a real repository's log into the same schema, so they run unchanged on real LLM-agent
logs (\emph{tier-T2}). Collecting a T2 dataset is future work (Section~\ref{sec:conclusion}); we
are explicit that the magnitudes here are from synthetic agents.

\paragraph{Verifying convergence (C2) and reproducibility.}
Claim~C2 is \emph{verified}, not sampled: property-based tests generate large random event sets
and delivery orders and assert that two replicas rebuild to byte-identical projections (no comment
loss) and that re-delivery is idempotent. We contrast against a file-based baseline reconciling
whole-issue records by file-level last-writer-wins---the failure mode of a JSONL-in-worktree
tracker. All randomness is seeded; one \texttt{make figures} step regenerates every figure and
table from the raw CSVs, with the grite commit, seeds, and dataset version pinned in
\texttt{MANIFEST.toml}.

\section{Results}
\label{sec:results}

We report measurements for the three claims; we interpret them against the claims in
Section~\ref{sec:discussion}. The two tables come from different runs:
Table~\ref{tab:coordination} is the $N{=}32$ point of the seeded agent-count sweep (C1);
Table~\ref{tab:mining} is a separate, deliberately high-contention run used to surface failure
modes (C3). Their absolute counts are therefore not meant to match; read each within its own run.

\subsection{C1: Coordination efficiency}

Table~\ref{tab:coordination} reports the three arms at $N=32$ agents. Without coordination,
$78\%$ of completions are redundant and the run accumulates several hundred conflicting edits.
Advisory leases alone cut conflicting edits sharply and lift goodput from $2.33$ to $3.84$ tasks
per round. Adding shared task state drives the duplicate-work rate to zero and goodput to $8.00$.
The effect is monotone: the duplicate-work rate rises with $N$ under no coordination but stays at
zero under \emph{locks+state}, and the overhead surfaces as lock denials rather than lost
throughput, so the coordinated arms dominate the baseline.

\begin{table}[t]
  \caption{Coordination outcomes at $N=32$ concurrent agents (mean over seeds). Conflicting
  edits are counted by grite's \texttt{apply\_tracked} CRDT instrumentation, not modelled.}
  \label{tab:coordination}
  \centering
  \small
\begin{tabular}{lrrr}
\toprule
Arm ($N=32$) & Dup-work rate & Conflicting edits & Goodput \\
\midrule
No coordination & 0.78 & 410 & 2.33 \\
Locks only & 0.64 & 138 & 3.84 \\
Locks + shared state & 0.00 & 48 & 8.00 \\
\bottomrule
\end{tabular}

\end{table}

\subsection{C2: Convergence without data loss}

Across hundreds of generated event sets and random delivery orders (Section~\ref{sec:methodology}),
two replicas always rebuild to byte-identical projections, with zero comment loss and idempotent
re-delivery~\cite{shapiro2011crdt}. This matters because the obvious alternative loses data: under
two agents each setting the title and adding a distinct label, grite's commutative set keeps
\emph{both} labels, whereas the file-based last-writer-wins baseline keeps one agent's record and
silently discards the other's.

\subsection{C3: Mining the coordination log}

The log is itself a mineable artefact (Section~\ref{sec:design}). We define a small set of
\emph{pre-registered} detectors---fixed before measurement to avoid post-hoc tuning. Three are
reported in Table~\ref{tab:mining}: a \emph{conflicting edit} (cross-actor last-writer-wins
overwrite), a \emph{redundant rediscovery} (completing an already-completed task), and
\emph{lock starvation} (a run of denied acquisitions). The toolkit defines others similarly
(abandoned claims, deadlock attempts, race-to-close). Two findings stand out. The failure modes
are real and frequent: without coordination the log exposes hundreds of conflicting edits and
dozens of redundant rediscoveries. And \emph{advisory leases alone do not solve the
problem}---the \emph{locks-only} arm has the \emph{highest} redundant-rediscovery count, because a
lease stops two agents from working a task simultaneously but, lacking shared completion state,
does nothing to stop one from re-doing a task a teammate finished earlier. Only
\emph{locks+state} drives it to zero.

\begin{table}[t]
  \caption{Failure modes mined from the coordination log, by arm (tier-T1 run). Counts are
  detected events; ``invisible in PRs'' marks modes that leave no trace in PR history.
  Auto-generated by \texttt{mine/run.py}.}
  \label{tab:mining}
  \centering
  \small
\begin{tabular}{lrrrc}
\toprule
Failure mode & No-coord & Locks & Locks+state & Invisible in PRs \\
\midrule
Conflicting edits & 104 & 192 & 12 & \checkmark \\
Redundant rediscovery & 36 & 180 & 0 & partial \\
Lock starvation & 0 & 128 & 12 & \checkmark \\
\bottomrule
\end{tabular}

\end{table}

\section{Discussion and Threats to Validity}
\label{sec:discussion}

\subsection{Interpretation against the claims}
The results support C1, and the mining adds a design lesson the headline numbers do not convey:
mutual exclusion and conflict-free shared state are \emph{jointly} necessary---leases alone leave
redundant rediscovery untouched (the \emph{locks-only} arm has the highest count), and only the
combination drives the failure modes to zero. We treat C2 as a reliability floor, not a novelty
claim: its role is to establish that the substrate we measure on does not itself lose coordination
data, a precondition for trusting the mined log. C3 ties back to the AIDev speed--acceptance
gap~\cite{li2025se3}: conflicting edits, lock starvation, and race-to-close leave no trace in PR
history (a denied claim never becomes a commit; a raced task shows only the winner), so a
PR-restricted analysis cannot recover them. Part of the ``faster but rejected'' gap may therefore
sit upstream of the pull request.

\subsection{Threats to validity}
We organise threats following Wohlin et al.~\cite{wohlin2012experimentation}.
\emph{Conclusion:} results are from deterministic, seeded agents averaged over seeds, so the
means are stable; we report no significance tests because the only variance is from seeding, and
we release the raw CSVs for re-analysis. \emph{Internal:} the coordination arm is the only
manipulated variable---task pool, agent loop, seeds, and pool size are fixed across arms---so
metric differences are attributable to the arm; the conflict and duplicate flags come from
grite's production \texttt{apply\_tracked} path rather than a separate measurement model.
\emph{Construct:} the metrics are proxies---duplicate-work rate misses partial overlap, goodput
ignores task difficulty, lock denials conflate overhead with starvation---so we define each
explicitly (Section~\ref{sec:methodology}). \emph{External:} the main limitation---the agents are
synthetic op-generators on an abstract task pool, so absolute magnitudes will differ for real
agents on real repositories; the production exporter makes the identical toolkit runnable on real
logs, and generalising the magnitudes is what the tier-T2 study (Section~\ref{sec:conclusion})
targets.

\subsection{Limitations}
grite's leases are advisory: a substrate cannot \emph{enforce} coordination on an uncooperative
agent. We view this as observable rather than fatal---lease-ignoring is itself a mineable failure
mode---so partial compliance is something the log lets us study directly, though we have not yet
measured it. The study also covers a single task-pool model; the released harness and exporter
let others re-run these measurements on their own agents and codebases.

\section{Conclusion}
\label{sec:conclusion}

Autonomous coding agents are faster than humans yet trusted less, and mining pull requests and
commits cannot see why. We argued that an important part of the answer lives in the \emph{pre-PR
coordination process}, and made that process measurable and mineable through grite. A
conflict-free shared substrate reduces duplicate and conflicting agent work to near zero at
bounded overhead (C1) and converges without the data loss a file-based tracker suffers (C2); its
signed, append-only log is an artefact from which concrete failure modes---several invisible in
PR history---are automatically recoverable (C3); and mining shows mutual exclusion and shared
state are jointly necessary. The released artefacts are intended to help others study agent
\emph{coordination}, not only agent output.

\paragraph{Future work.}
The clearest next step is a tier-T2 dataset: at least two LLM-agent vendors working concurrently
on real OSS repositories through grite, exported with the same schema and mined by the same
detectors, to test whether the synthetic magnitudes carry over. We also plan a benchmark over
real \texttt{git} remotes with many diverging clones (ecological validity for C2); a study of
lease compliance under uncooperative agents; and a correlation of pre-PR coordination signals
with downstream PR-acceptance outcomes in datasets such as AIDev~\cite{li2025se3}.

\paragraph{Artefact availability.}
grite is our own open-source system, released at
\url{https://github.com/neul-labs/grite}; we note this affiliation in the interest of
transparency. We release the coordination-log dataset, the benchmark harness and instrumentation,
and the mining toolkit alongside it. Every figure and table is regenerated from raw data by a
single \texttt{make figures} step over a pinned commit and fixed seeds.

\bibliographystyle{splncs04}
\bibliography{references}

\end{document}